\documentclass[english,onecolumn,prl,superscriptaddress]{revtex4-1}
\usepackage[T1]{fontenc}
\usepackage[latin9]{inputenc}
\usepackage{color}
\usepackage{stmaryrd}
\usepackage{babel}
\usepackage{units}
\usepackage{textcomp}
\usepackage{amsmath}
\usepackage{amssymb}
\usepackage{graphicx}
\usepackage{setspace}
\usepackage{cancel}
\usepackage{xcolor}
\usepackage{esint}
\usepackage[unicode=true,pdfusetitle,bookmarks=true,bookmarksnumbered=false,bookmarksopen=false, breaklinks=false,pdfborder={0 0 1},backref=false,colorlinks=true]
 {hyperref}
\hypersetup{linkcolor=blue,urlcolor=blue,citecolor=blue}
 \makeatletter

\begin{document}
\title{Exotic states of matter in an oscillatory driven liquid crystal cell}


\author{Marcel G. Clerc}
\affiliation{ Departamento de F\'isica and Millennium Institute for Research in Optics, FCFM, 
 Universidad de Chile, Casilla 487-3, Santiago, Chile.}
\author{ Michal Kowalczyk}
\affiliation{  Departamento de Ingenier\'ia Matem\'atica and Centro de Modelamiento Matem\'atico 
 (UMI 2807 CNRS), Universidad de Chile, Casilla 170 Correo 3, Santiago, Chile}
\author{Valeska Zambra}
\affiliation{ Departamento de F\'isica and Millennium Institute for Research in Optics, FCFM, 
 Universidad de Chile, Casilla 487-3, Santiago, Chile.}

\begin{abstract}
{\bf Abstract.} Matter under different equilibrium conditions of pressure and temperature exhibits different states such as
solid, liquid, gas, and plasma. 
Exotic states of matter, such as Bose-Einstein condensates, superfluidity, chiral magnets, 
superconductivity, and liquid crystalline blue phases are observed in thermodynamic equilibrium.
Rather than being a result of  an aggregation 
of matter, their emergence  is due  to a change of a topological state of the system.   Here
we investigate topological states of matter in a system with injection and dissipation 
of energy. In an experiment involving  a liquid crystal cell under the influence of a low-frequency oscillatory 
electric field, we observe a transition from non-vortex state to a state in which  vortices persist.
Depending on the period and the  type of the forcing, the vortices self-organise forming square lattices, 
glassy states, and disordered vortex structures. Based on a stochastic amplitude equation, 
we recognise the origin of the transition as the balance between stochastic creation and deterministic annihilation of vortices. 
 Our results show that the matter maintained 
out of equilibrium by means of the temporal modulation of parameters can exhibit exotic states.
\end{abstract}

\maketitle

\section{Introduction}
Solid, liquid, gas, and plasma  are different states of the matter\cite{Goodstein}  distinguished from each other by mechanical, optical, 
and other properties. 
Other examples of   states of aggregation of matter include  glassy and liquid crystal states.
Still different are exotic states 
such as Bose-Einstein condensates \cite{Pethick}, superfluidity \cite{Tsuneto},  superconductivity \cite{Tinkham},
chiral magnets \cite{Skyrmion2009}, and liquid crystalline blue phases \cite{Gennes} that are a topological state
 rather than an aggregation of matter. The topological transitions of the matter
were discovered at the beginning of the 70s by Berezinskii \cite{Berezinskii} and Kosterlitz and Thouless \cite{Kosterlitz}, 
who showed that a low dimensional system described by a physical vector order parameter in thermodynamic equilibrium 
undergoes  a transition from a homogeneous state without vorticity to a state  in which vorticity persists.
In the homogeneous state all vectors are unidirectionally ordered but under suitable conditions they realign
forming  regions where both their orientations and magnitudes vary.  
Because of topological constraints at some  isolated points called vortices \cite{Pismen}  the vector field  vanishes and  the vector phase becomes undefined. 
The winding number (topological charge) is introduced to characterise the physical vector field around a vortex \cite{Pismen}. 
This number is an integer representing the total number of times that the vector field winds around the origin while varying along a closed, counterclockwise oriented curve  around the singular point.
Topological stability of the system implies that the total winding number of the system must be preserved which means that the vortices are created or annihilated by pairs of opposite topological  charges.
Vortices creations and and annihilation process are, respectively, due to  thermal fluctuations and free energy 
minimisation \cite{Dierking2005,Barboza2013}, hence at a critical temperature at which they are balanced the systems 
undergoes a  topological transition \cite{Berezinskii, Kosterlitz}.
Exotic states of matter such as Bose-Einstein, superconductivity, chiral magnets, and superfluidity are usually observed at low temperatures, however  liquid crystalline blue phases have been observed at room temperature \cite{Coles2005}.

\begin{figure}
\begin{center}
\includegraphics[width= 12 cm]{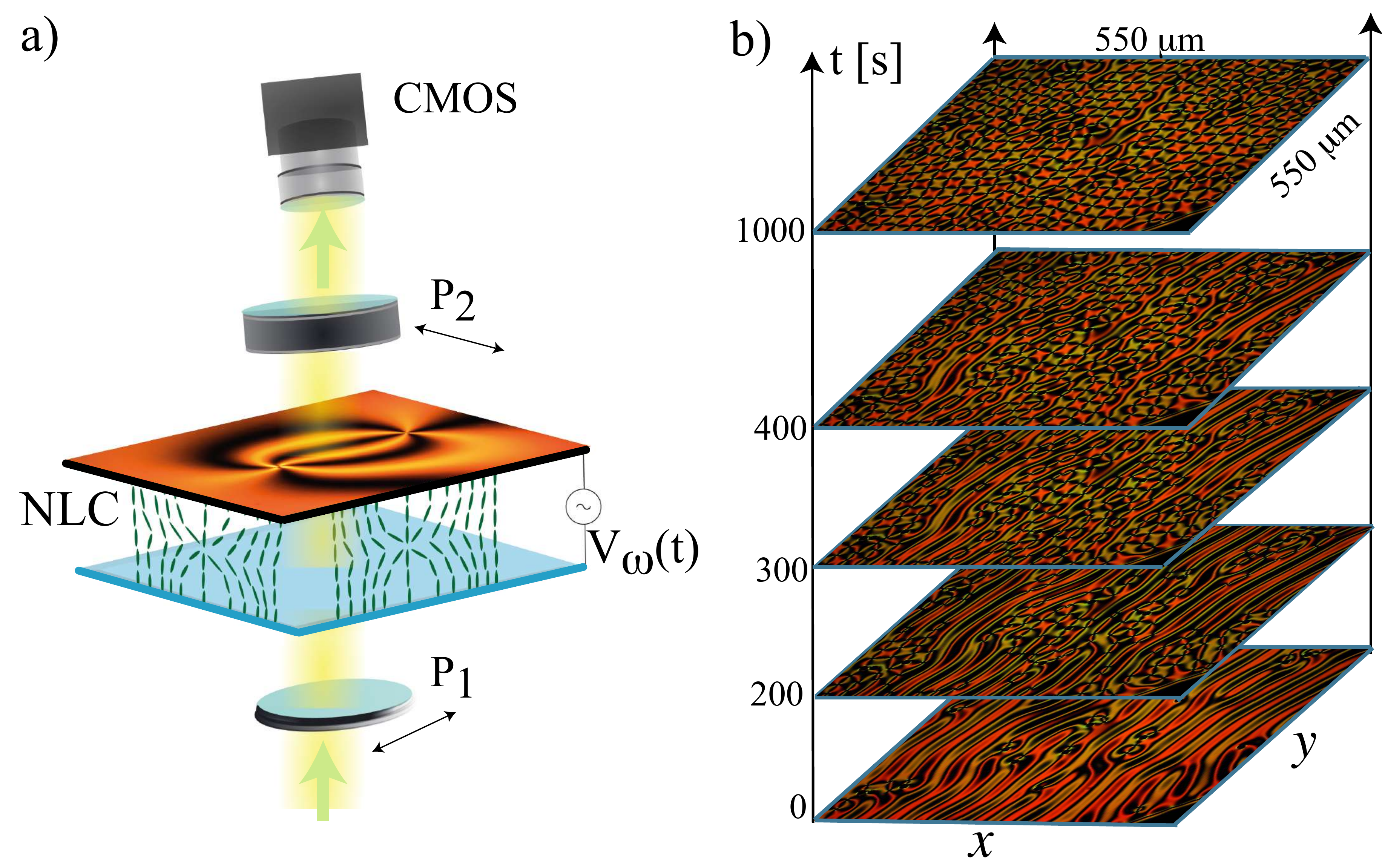}
\caption{Liquid crystal cell under a temporarily modulated potential exhibits creation and self-organisation 
of vortices. (a) Schematic representation of the experimental setup. 
Liquid crystal cell (NLC) with homeotropic anchoring is illuminated by white light between two crossed polarisers ($P_1$ and $P_2$).
The horizontal snapshot shows a pair of vortices with opposite charges.
(b) The temporal sequence of snapshots in the region of self-organised vortices, at frequency~0.335~Hz and voltage amplitude $13.5$~Vpp. From experimental snapshots, both figures were created using Inkscape 1.0.}
\end{center}
\label{Fig1-Setup}
\end{figure}

An ideal material to investigate vortex dynamics  are liquid crystals in thin films \cite{Gennes,Chandrasekhar}. 
One of the most studied vortices are the so-called umbilical defects or disclination lines \cite{Gennes,Chandrasekhar,Rapini1973}.
In thermodynamic equilibrium and homogeneous media, the vortices tend to annihilate by pairs to minimise the free energy of the system. The above dynamics can be modified by means of incorporation of inhomogeneities, which can attract and trap 
umbilical defects \cite{Kim2020,Zambra2020}. Properly distributed inhomogeneities may permit the formation 
of topology lattice \cite{Kim2020}. Likewise, considering inhomogeneous anchoring allows attracting and trapping 
umbilical defects and creating vortex lattices \cite{Marrucci2006,Murray2014}.
A similar effect can be achieved by the introduction of inhomogeneous electrodes \cite{Brasselet2014, Orihara2016,Salamon2018,Murray2020}.
The combined use of magnets and uniform electric field can induce umbilical defects and lattices \cite{Pieranski2013}.
The vortex lattices describe above are induced by the combination of the forcing and inhomogeneities.
However, the emergence of spontaneous topological lattices has also been achieved 
by means of thermal gradients \cite{Pieranski1073} or by doping with ionic impurity \cite{Orihara2016}, 
which induces charge motions. This is known as the Carr-Helfrichh mechanism \cite{Gennes}.
This article aims to study exotic states of matter with injection and dissipation 
of energy. This type of physical context usually is denominated  as out of equilibrium systems \cite{Prigogine,Haken}.
Based on an experiment involving a nematic liquid crystal cell under the influence of a low-frequency oscillatory 
electric field, we observe a transition from non-vortex state to a state in which vortices persist. 
Depending on  the frequency and the type of the forcing, the vortices self-organise forming 
square lattices, glassy states, and disordered vortex structures. 
Theoretically, a stochastic amplitude equation allows us to reveal  
the origin of the transition in terms of the balance between stochastic creation and deterministic annihilation of vortices.

\section{Results}
{\bf Experimental observations of a topological transition in a driven liquid crystal cell.} 
Liquid crystals are composed of rod-like organic molecules \cite{Gennes, Chandrasekhar,Rapini1973} which, 
as a result of intermolecular interaction, for specific temperature ranges   
are arranged to have a similar molecular orientation. 
This results in a strong anisotropy of all their physical properties, especially optical characteristics \cite{Blinov}. 
The configuration of lowest energy is reached when all rod-like molecules 
are aligned along one averaged direction,  orientational order without  a positional one, 
denoted by the director vector ${\bf n}$ \cite{Gennes,Chandrasekhar,Blinov}.
This state is usually called the nematic phase.
In the case of  a thin film with negative dielectric anisotropy 
and molecular anchoring perpendicular to the walls of the sample, application of  an electric field 
in the vertical direction leads to the appearance of  vortices, umbilic defects or disclination lines \cite{Gennes,Chandrasekhar,Rapini1973}.
Figure~1b shows the spatiotemporal evolution of  vortex arrangements experimentally observed by applying  a
voltage  $V(t)=V\sin(2\pi f t)$ with a given frequency $f$, i.e, harmonic voltage signal.
To avoid charges accumulation effects in the thin film (capacity effects), a high 
frequency oscillatory electric field (kHz) is usually used.
Under these conditions in a homogeneous liquid crystal cell
the emergence of gas of disordered vortices is followed  
by the subsequent annihilation  by pairs,
and terminates  in a homogenous, non-vortex state \cite{Dierking2005,Barboza2013,Chandrasekhar}. 
Thus the vortices are a transient phenomenon.
Surprisingly, when  the frequency of the electric field  that we applied to the homogeneous liquid crystal cell decreases 
to fractions of Hz starting from a critical value of the frequency, the system exhibits 
a topological transition after which  the annihilation and creation are balanced, 
and  the  vortices persist (see video 1 in supplementary materials).
Hence, the bifurcation parameter of this transition is the frequency  $f$ of the drive voltage.
Figure 2 shows 
the average number of vortices as a function of frequency 
counted stroboscopically in each oscillation cycle with
 the standard deviation  determined along the way.
  This transition is obtained by considering a sawtooth signal for the voltage applied to the sample.
From this chart, one deduces that the transition is of continuous 
nature (supercritical bifurcation) and that there is a critical frequency $f_c$ 
from which the number of vortices in average  becomes permanent over time (frequency $<f_c$).
We note that as the frequency decreases the number of vortices increases to 
a particular critical value and subsequently decreases monotonically  until it vanishes 
at low frequencies, which is a manifestation of a sort of resonance for the process of creation 
and destruction of topological defects. 
Notice that periodically driven  voltage only induces umbilical defects, no other defects are observed.
The application of a low-frequency electric field induces charge movements due to the 
weak anisotropic conductivity of the liquid crystal \cite{Gennes}. The accumulation of charges 
can induce a molecular reorientation, Carr-Helfrich mechanism \cite{Gennes}, 
which in turn modifies the interaction between umbilical defects and can even generate a lattice arrangement of
them \cite{Pieranski1073}.

Using a thermal control microscope stage, the temperature of the liquid crystal sample 
can be changed and controlled adequately. 
When the temperature at which the experiments are made is varied, 
we observe that critical frequency transition $f_c$ grows monotonically with it as illustrated in Figure~2c.
The tendency to increase the transition frequency at higher temperatures is due to the increasing  
the rate of vortex creation (fluctuations), while the process of vortices annihilation remains unchanged (deterministic).
Therefore, the topological transition induced by temporal voltage modulation is 
observed throughout the mesophase stability range of the nematic liquid crystal under study.

\begin{figure}
\begin{center}
\includegraphics[width= 14 cm]{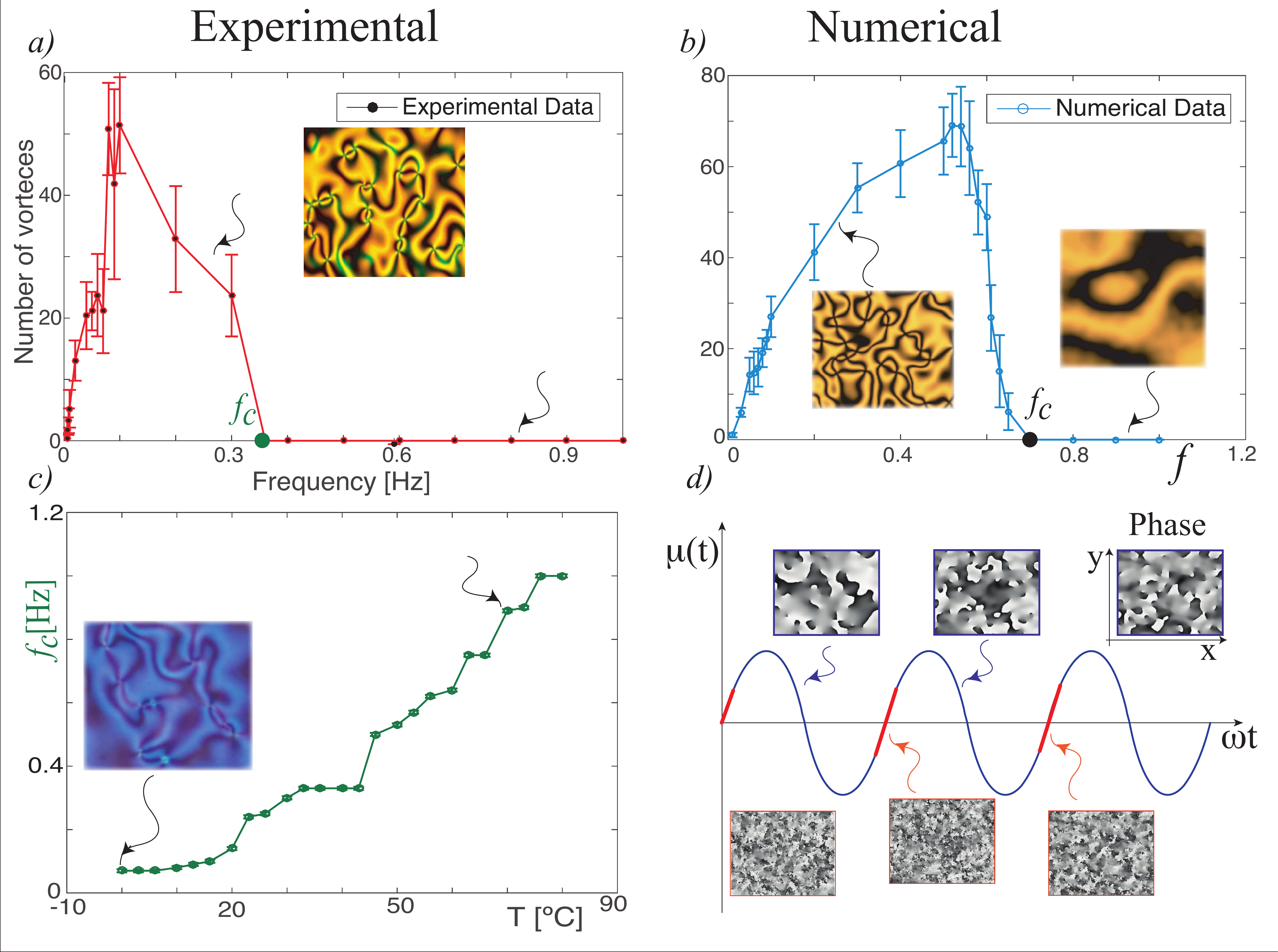}
\caption{Bifurcation diagram of topological transition out of equilibrium 
(a) experimental and (b) numerical using model Eq.~(1).
The experimental bifurcation diagram is obtained with a sawtooth forcing with 
 a fixed  amplitude voltage $15$~Vpp.
(c) Critical frequency $f_c(T)$ as a function of temperature  with a fixed  amplitude voltage $15$~Vpp.
The insets account for the respective snapshots in the different regions.
(d) Evolution of the temporal bifurcation parameter  $\mu(t)$ and characterisation of 
the regimes of creation (red curve) and interaction (blue curve) of vortices. 
Inserts show the phase obtained numerically in the different creation and interaction regimes.
From experimental snapshots, both figures were created using Inkscape 1.0.
}
\end{center}
\label{Fig1-Setup}
\end{figure}

{\bf Theoretical description of the topological transition.} To understand the origin of this topological transition out of equilibrium, we consider a prototype model, the Ginzburg-Landau equation
\cite{AransonKramer2002},
that describes the emergence of topological defects in fluids, superfluids, superconductors, liquid crystals, 
chiral magnets, 
fluidised anisotropic granular matter, and magnetic media \cite{AransonKramer2002,Pismen}.  
The real Ginzburg-Landau equation describe the pattern formation in anisotropic media \cite{PeschKramer1986}. 
Likewise, this model  describes  vortex solutions in nematic liquid crystal layers with external electric or magnetic forcing 
and  homeotropic boundary conditions \cite{Frisch1994,CoulletPlaza1994,Frisch1995,ClercKowalczyk2014},
and  the formation of spiral waves in a nematic liquid crystal subjected to a rotating magnetic \cite{Frisch1994,Frisch1995}
 or electric field \cite{CoulletPlaza1994}.
 Note that this Ginzburg-Landau equation with real coefficients is derived from the elastic theory 
 of liquid crystals \cite{Frisch1994,CoulletPlaza1994,Frisch1995,ClercKowalczyk2014,ClercVidal2015}.
 The order parameter accounts for the balance between the elastic and electric force.
Besides, this model  describes  the process of interaction and 
 annihilation of vortices at constant electric field and temperature \cite{Barboza2013}. To account for the additional  ingredients of the observed 
topological transition (cf. Fig.~2), we must incorporate the oscillatory nature of the electrical 
voltage applied to the liquid crystal sample and include the inherent fluctuations due to temperature.
This leads to  the stochastic Ginzburg-Landau equation with oscillatory coefficients, that is,
\begin{equation}
\partial_t A=\left[\mu_0+\gamma \cos(2 \pi  f t)\right] A-|A|^2A+\nabla^2 A+\sqrt{T}\zeta(\vec{r},t)
,\label{Eq-GL}
\end{equation}
where $A(\vec{r},t)$ is a complex order parameter, $t$ and $\vec{r}$ describe time and the transversal coordinate
vector that characterises the thin film,  $\mu_o$ is the uniform bifurcation parameter, $\gamma$ and $f$ 
are the amplitude and the frequency of the forcing, respectively, which account for the oscillatory electric field. 
The function  $\mu(t) =\mu_0+\gamma \cos( 2 \pi f t)$
is the temporal modulated bifurcation parameter.  
By $\nabla^2$ we denote the Laplace operator. The constant $T$ accounts for the thermal intensity and $\zeta(\vec{r},t)$ is a 
spatiotemporal white noise of zero mean value,  $\langle \zeta(\vec{r},t) \rangle=0$,
 and no spatial or temporal memory. Namely, the stochastic term has the spatiotemporal correlation 
 $\langle \zeta(\vec{r},t) \zeta(\vec{r}',t') \rangle=\delta(\vec{r}-\vec{r}')\delta(t-t')$ where $\delta$ are Dirac delta functions.  

\begin{figure}
\begin{center}
\includegraphics[width= 14 cm]{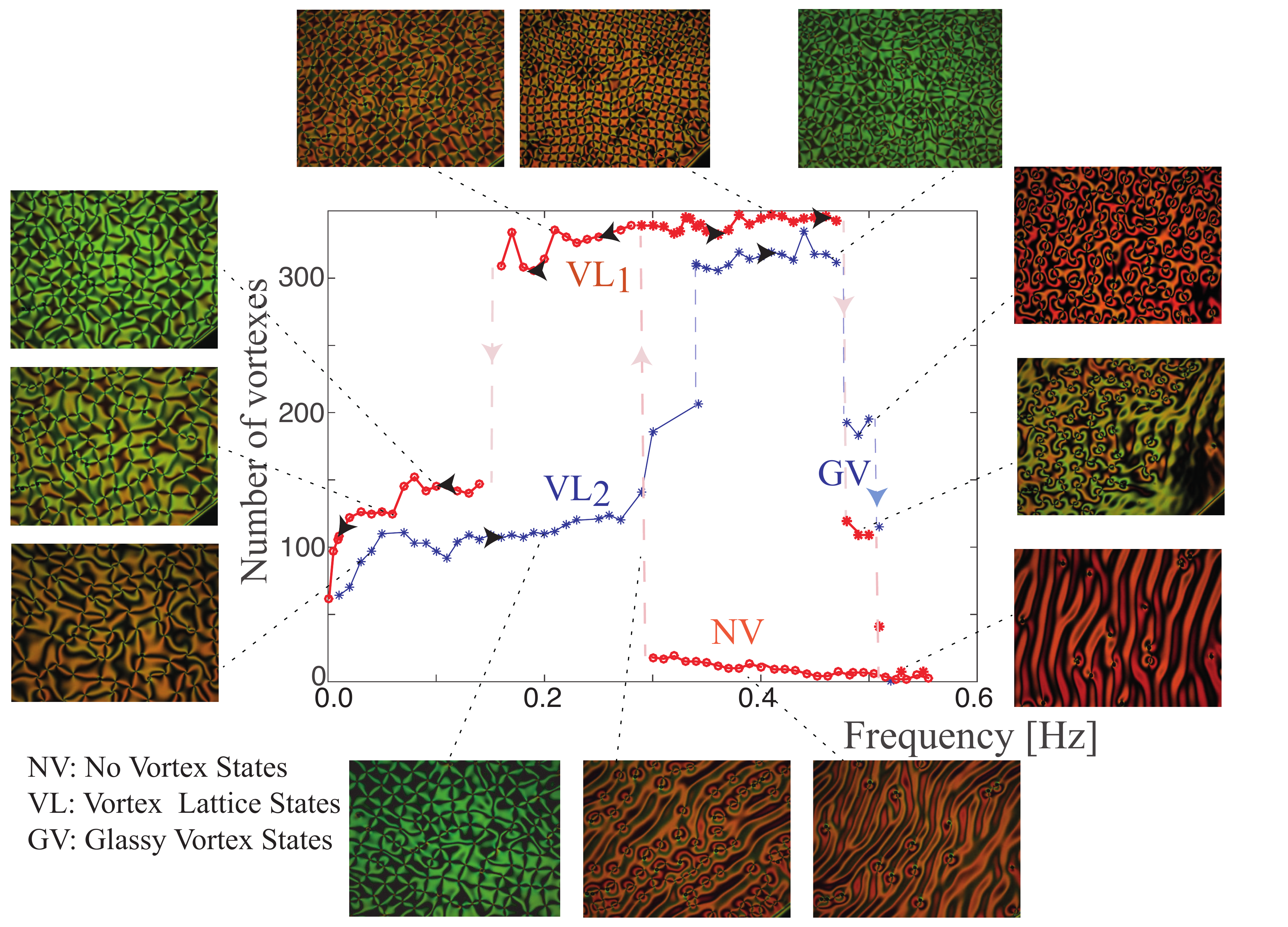}
\caption{Experimental bifurcation diagram of topological transition out of equilibrium under harmonic forcing.
The liquid crystal cell exhibits three states: 
non-vortex (NV), vortex lattice (VL) and glassy vortex (GV) states.
The arrows indicate the direction of increase or decrease of the voltage.
The insets show snapshots in the respective parameter ranges.
From experimental snapshots, figures were created using Inkscape 1.0.
}
\end{center}
\label{Fig3-Setup}
\end{figure}

 In the high-frequency regime, $f \to \infty$, 
 this model becomes the Ginzburg-Landau equation with real coefficients. 
 This equation is characterised 
 by a constant effective bifurcation parameter $\mu_0+3\gamma^2/2(2\pi f)^2$ obtained through the rapid 
 oscillation method \cite{LandauLifshitzCM}. 
 In this limit the vortices do not persist
 and the annihilation of the defects of opposite charges dominates their creation \cite{Pismen,Barboza2013}, 
 since the system tries to optimise the effective free energy.
 Figure~2b shows this happening for frequency values up to order one.
  In this regimen, for large enough temporary evolution, the number of vortices on average is zero.
 By decreasing the frequency further  to a critical value $f_c$, the average number of vortices stabilises over time.
  The topological transition obtained numerically using Eq.~(\ref{Eq-GL}) 
 has  a qualitative behaviour similar to that observed experimentally, see top panels in Fig.~2.
Notice that as the frequency decreases ($f<f_c$) the number of vortices increases to 
a particular critical value and subsequently decreases monotonically  until it vanishes 
at low frequencies, which manifests an excellent qualitative agreement with the experimental observations.
Hence, experimentally and numerically a sort of resonance is observed for the process of creation 
and destruction of topological defects.

 The simulation allows us to identify the location of the  vortices through $\pm2\pi$ jumps of the phase of the amplitude.
Comparing  the evolution of the system and the profile of  the bifurcation parameter function $\mu(t)$ two characteristic regions are identified. 
Namely, a creation and annihilation region.
Creation of vortices occurs in  the intervals of time where  $\mu(t)$ is small and growing
 (red curve in Fig.~2d), these vortices later interact 
even when $\mu(t)<0$  (blue curve in Fig.~2d). 
The region of creation and annihilation are govern by stochastic fluctuations and deterministic evolution, respectively.
The vortex creation time interval  decreases as  the forcing frequency increases and for high frequencies the creation process is inefficient.
Hence, the persistence  of vortices is a consequence of the balance between the processes of creation 
(stochastic) and their interaction (deterministic).

{\bf Topological transition with harmonic driven forcing.} In experiments  we have implemented various  types of periodic forcing 
among them harmonic, sawtooth, or square profiles and we have found, somewhat  unexpectedly, different 
types of responses resulting in diverse transitions.
 As we have mentioned, low-frequency voltages can induce charge movements that, 
in turn, induce molecular reorientation, Carr-Helfrich mechanism \cite{Gennes}. 
Hence, different types of driven voltages can induce different charge motions.
In the case of a 
square profile  signal, we have observed a continuous or supercritical topological transition (see Fig.~1a). 
Changing to  a harmonic signal, we have detected  a discontinuous transition with the non-vortex  state being  replaced by a vortex lattice with a square crystalline structure.
Figure 3 shows a square vortex lattice and its respective bifurcation diagram corresponding to the out of equilibrium counterpart of Abrikosov lattice \cite{Abrikosov,Ketterle}.
The vortex lattice 
in not hexagonal like the one  of Abrikosov as a consequence of the asymmetry between the opposite charges \cite{ClercKowalczyk2014}. 
The model Eq.~(1) only accounts for the topological transition from disordered vortices to non-vortex state. 
The origin of these square vortex lattices is probably associated with the coupling of elastic deformations and fluid modes. 
To account qualitatively of this coupling, we include in the model Eq.~(1) inertia  and anisotropic effects, 
that is, a second temporal derivative of amplitude A. Simulations of this model show 
the emergence of a square lattice, as seen in Fig.~4.
The Ginzburg-Landau Eq.~(\ref{Eq-GL}) is a model infer close to the reorientational transition \cite{Frisch1995,ClercVidal2015}. 
Its derivation is based on the assumption of slowly varying amplitude; 
however, when the system is periodically forced, the first and second temporal variations can be of the same order. 
The inertia term phenomenologically accounts for the effects of movements of charges and liquid crystal inside the cell.

\begin{figure}
\begin{center}
\includegraphics[width= 12 cm]{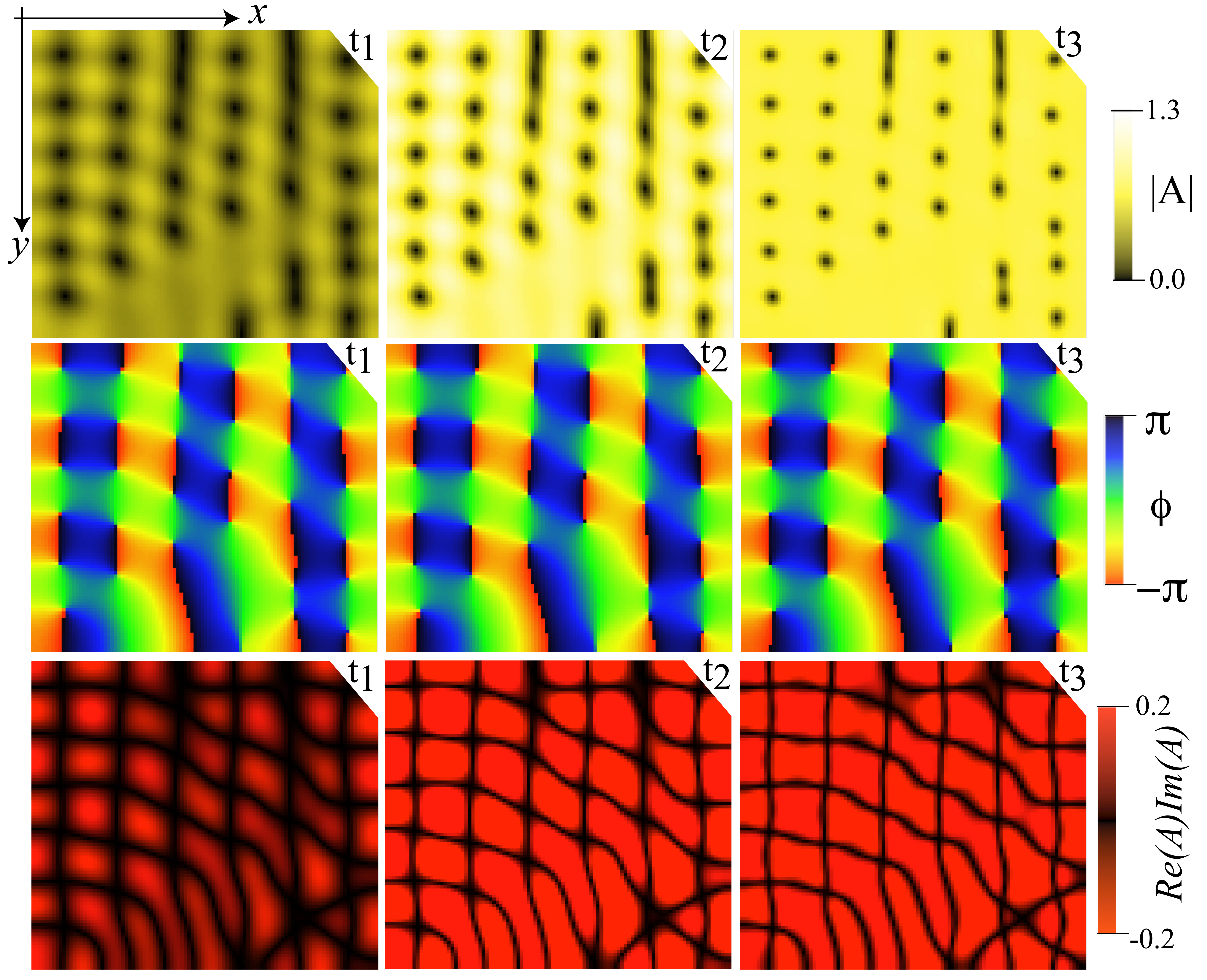}
\caption{Numerical square vortex lattice. 
Temporal sequence ($t_1 <t_2 <t_3$) of the amplitude module $|A|$, phase $\phi=arctan[Im(A)/Re(A)]$, and polarisation field
$Re(A)Im(A)$  of model Eq.~(2) 
with inertia and anisotropic coupling over a period,
by $\mu_0=0.6$, $T=0.03$, $\lambda=1.4$, $\delta=0.3$, $\gamma=3$, and $f=0.1$. 
From numerical simulations, figure was created using Inkscape 1.0.
}
\end{center}
\label{Fig4-textures}
\end{figure}
 
When decreasing the frequency, the square lattice undergoes a subcritical bifurcation  
leading to  a square lattice of higher wavelength (see Fig.~3 and supplementary video 2).
Increasing the frequency  further the square lattice  transitions to a glassy state (cf. Fig.~3 and supplementary video 3), 
in which the  vortex structure
does not have a  precise unit cell. For even higher frequencies the system returns to the non-vortex state.
Figure 3 summarises the complexity of the topological transitions in the 
liquid crystal cell maintained out of equilibrium at room temperature.
We speculate that the origin of the periodic structures we have discovered may be 
associated with the interaction between the vortices or the excitation of stationary waves \cite{Migara},  
however, precise  understanding of this  is an open problem. 

\section{Discussion}

Topological defects in liquid crystals are natural elements used for the generation of optical vortices
\cite{Marrucci2006,Brasselet2009,Barboza2012,Chen2018,Salamon2018}. As a matter of fact, optical vortices have attracted attention 
for their diverse photonic applications ranging from optical tweezers \cite{Grier2003,Padgett2011}, 
quantum computation \cite{Arnaut2000}, enhancement of astronomical images \cite{Tamburini2006}.
In all these applications, optical vortex lattices are always involved and required
 \cite{Wang2012, BarbozaVidal2013,Lei2015,Stoyanov2018}.
These vortex lattices require sophisticated and complex experimental setup.
 Instead, he vortex lattice that we observe emerge spontaneously in simple  liquid crystal cells that do 
 not require a complex structure of electrodes, inhomogeneities, applications 
 of thermal gradients, combined forcing of electric and magnetic fields, or photosensitive walls.

In conclusion, we have shown that exotic states of matter with injection and dissipation 
of energy. In a nematic liquid crystal cell under the influence of a low frequency oscillatory 
electric field, we have observe transitions from non-vortex state to a state in which vortices persist.
Depending on the frequency and  type of the forcing (harmonic, sawtooth, or square profiles), 
the vortices self-organise forming square lattices, 
glassy states, and disordered vortex structures. 
Because the phenomenon reported  here is qualitativelly well described by a universal model Eq.~(\ref{Eq-GL}), 
we expect that any temporally modulated  vectorial field system of low dimensionality 
can exhibit topological transitions out of equilibrium.
The characterisation of the  critical  frequency and voltage 
as a function of liquid crystal features and cell configuration  is an open question. Work in this direction is in progress.
Furthermore, these findings  could be a starting point for understanding and controlling 
the exotic states of matter  out of equilibrium by means of the temporal modulation of parameters.
Because vortex lattices emerge spontaneously in single cells subjected to alternative low-frequency voltages, 
it opens up the possibility of new and fresh applications of the generation of optical vortices.

\bigskip
{\bf Methods:}
{\bf Experimental description of the setup.}
Figure 1a shows a schematic representation  of the experimental setup. 
It consists of a liquid crystal cell composed of  two glass slabs  with  600~mm$^2$ of cross-section separated by a distance of 15~$\mu$m, 
 a thin film of a transparent conductor, indium tin oxide (ITO), and a thin film of transparent polyimide 
 that has been deposited on each of the interior walls. 
Transparent conductors are used as electrodes. By rubbing process,
microscopic grooves are generated in the polyimide layer, 
allowing the liquid crystal molecules anchoring orthogonally to the surfaces, 
homeotropic anchoring. This cell  5B100A150UT180 manufactured by Instec,
contains glass beads as spacers. 
It is filled by capillarity with BYVA-01-5G (Instec) nematic liquid crystal that has negative anisotropy,
$\epsilon_a=-4.89$ at room temperature. 
An external electric field is applied in the vertical direction (z-axis) using a sinusoidal 
sawtooth, or square voltage with amplitude 
15 Vpp with low frequency. This voltage is produced by a function generator (Agilent 33521A) 
with a high voltage amplifier (Tabor Electronics 9200).
The imaging system used is an Olympus BX51  microscope equipped with linear cross polarisers. 
The light from the microscope condenser illuminates the cell mounted on the microscope stage, 
and a CMOS camera (Thorlabs  DCC1645C) is used to capture images. 
For studying thermal effects we used Leica DM2700 P microscope  equipped with  LTS420 hot stage.

{\bf Numerical Simulations.}
Numerical simulations of model Eq.~(1) were implemented using a finite differences 
code with Runge-Kutta order-4 algorithm,  with a 200$\times$200 points grid, spacing $dx=0.5$, 
and temporal increment $dt=0.02$. 
Numerical simulations are performed with periodic boundary conditions and with an initial condition $A = 0$.
The stochastic noise $\zeta(\vec{r},t)$ is generated through the Box-Muller transform 
of a uniform random number generator.  Equation~(1) with inertia  and anisotropic effects reads 
\begin{equation}
\partial_{tt} A+\lambda \partial_t A=\left[\mu_0+\gamma \cos(2 \pi  f t)\right] A-|A|^2A+\nabla^2 A
+\delta \partial_{\eta,\eta}\bar{A}+\sqrt{T}\zeta(\vec{r},t), 
\end{equation}
where $\lambda$ accounts for the rotational viscosity, $\delta$ stands for the difference of elastic constants \cite{Frisch1994,CoulletPlaza1994,Frisch1995,ClercKowalczyk2014}, the operator 
$\partial_{\eta,\eta}=\partial_{xx}-\partial_{yy}+2i\partial_{xy}$ describes the asymmetric coupling,
and $\bar{A}$ is the complex conjugate of $A$.
The results presented in figure~4  consider the same algorithm, boundary and initial conditions used in equation~(1).



\bigskip
 {\bf Acknowledgements:}
The authors thank M. Morel and G. Gonzalez for fruitful discussions.
The authors acknowledges the financial support of the Millennium Institute for Research in Optics (MIRO), Fondecyt
project 1180903 and 1170164, and Fondo Basal CMM-Chile.

\bigskip
 {\bf Author contributions} M.G.C. and V.Z. conceived the experiments.
 V.Z. performed the experiments and analysed the data. M.G.C. and V.Z. 
performed the numerical computations. M.G.C., M.K., and V.Z. worked on the theoretical description.
 M.G.C. wrote the first draft  of the article. All authors contributed to
the overall scientific interpretation and edited the manuscript.
 
 \bigskip
 {\bf  Competing interests:} 
 The authors declare that they have no
competing financial interests.

\bigskip
 {\bf Correspondence:} Correspondence and requests for materials
should be addressed to M.G.C.~(email: marcel@dfi.uchile.cl), https://orcid.org/0000-0002-8006-0729.

\end{document}